\documentclass[twocolumn,showkeys,showpacs,preprintnumbers,prd,superscriptaddress,nofootinbib]{revtex4-1}

\usepackage{graphicx}
\usepackage{epsf}
\usepackage{bm}
\usepackage{amsmath}
\usepackage{amsfonts}
\usepackage{amssymb}
\usepackage{epstopdf}
\usepackage{natbib}
\usepackage{hyperref}
\usepackage{color}
\usepackage{verbatim}
\usepackage{multirow}
\usepackage{bm}
\usepackage{hyperref}

\begin{document}

\title{BAO signatures in the 2-point angular correlations and the Hubble tension}
\author{Rafael C. Nunes}
\email{rafadcnunes@gmail.com}
\affiliation{Divis\~ao de Astrof\'isica, Instituto Nacional de Pesquisas Espaciais, Avenida dos Astronautas 1758, S\~ao Jos\'e dos Campos, 12227-010, SP, Brazil}

\author{Armando Bernui}
\email{bernui@on.br}
\affiliation{Observat\'orio Nacional, 
Rua General Jos\'e Cristino 77, S\~ao Crist\'ov\~ao, 20921-400, Rio de Janeiro, RJ, Brazil}

\begin{abstract}
An observational tension on estimates of the Hubble parameter, $H_0$, using early and late Universe information, is being of intense discussion in the literature. 
Additionally, it is of great importance to measure $H_0$ independently of CMB data and local 
distance ladder method. 
In this sense, we analyze 15 measurements of the transversal BAO scale, $\theta_{\rm BAO}$, 
obtained in a weakly model-dependent approach, in combination with other data sets obtained 
in a model-independent way, namely, Big Bang Nucleosynthesis (BBN) information, 6 
gravitationally lensed quasars with measured time delays by the H0LiCOW team, and 
measures of cosmic chronometers (CC). 
We find $H_0  = 74.88_{-2.1}^{+1.9}$ km s${}^{-1}$ Mpc${}^{-1}$ and 
$H_0 = 72.06_{-1.3}^{+1.2}$ km s${}^{-1}$ Mpc${}^{-1}$  from $\theta_{BAO}$+BBN+H0LiCOW 
and $\theta_{BAO}$+BBN+CC, respectively, in fully accordance with local measurements. 
Moreover, we estimate the sound horizon at drag epoch, $r_{\rm d}$, independent of CMB 
data, and find $r_{\rm d}=144.1_{-5.5}^{+5.3}$ Mpc (from $\theta_{BAO}$+BBN+H0LiCOW) 
and $r_{\rm d} =150.4_{-3.3}^{+2.7}$ Mpc (from $\theta_{BAO}$+BBN+CC). 
In a second round of analysis, we test how the presence of a possible spatial curvature, 
$\Omega_k$, can influence the main results. 
We compare our constraints on $H_0$ and $r_{\rm d}$ with other reported values. 
Our results show that it is possible to use a robust compilation of transversal BAO data, 
$\theta_{BAO}$, jointly with other model-independent measurements, in such a way that the  tension on the Hubble parameter can be alleviated.
\end{abstract}



\maketitle
\section{Introduction}\label{Introduction}

The standard cosmological model, the flat $\Lambda$CDM, based on general relativity 
theory plus a positive cosmological constant and dark matter, has been able to explain 
accurately the most diverse observations made in the past two decades. 
Despite that, as new astronomical observations improve, in precision and in the diversity of 
cosmic tracers, arises a possible inability to explain within the standard paradigm 
quantitatively different measurements, and this is putting the $\Lambda$CDM cosmology in 
a crossroads. 
The most notable issue is the current tension on the Hubble parameter $H_0$. 
Assuming the $\Lambda$CDM scenario, Planck-CMB data analysis provides 
$H_0 = 67.4 \pm 0.5$ km s$^{-1}$Mpc$^{-1}$ \cite{Planck2018}, while a model-independent 
local measurement from Hubble Space Telescope observations of 70 long-period Cepheids 
in the Large Magellanic Cloud results 
$H_0= 74.03 \pm 1.42$ km s$^{-1}$Mpc$^{-1}$ \cite{R19}. 
These estimates are in $4.4\,\sigma$ tension. 
Additionally, a combination of time-delay cosmography from H0LiCOW lenses and the 
distance ladder results is at $5.2\,\sigma$ tension with CMB constraints \cite{H0LiCOW}. 
Another accurate independent measure was carried out in \cite{Freedman}, from Tip of 
the Red Giant Branch, showing $H_0 = 69.8 \pm 1.1$ km s$^{-1}$Mpc$^{-1}$. 
Other recent analysis also put in crisis the $\Lambda$CDM 
model~\cite{D_1,D_2,D_3,D_4,D_5,D_6}. 
In addition to this disagreement with diverse observations, it is important to remember that 
the  cosmological constant suffers from some theoretical problems~\cite{DE_01,DE_02} 
that motivates alternative scenarios that could, at the same time, explain the observational 
data and have some theoretical appeal. 
This stimulated recent discussions about whether a new physics beyond the 
standard cosmological model can solve the $H_0$ tension 
\cite{H0_0,H0_1,H0_2,H0_3,H0_4,H0_5,H0_6,H0_7,H0_8,H0_9,H0_10,H0_11,H0_12,H0_13,H0_14,H0_15}.

Other less noticed --but not less important-- issue concerns the standard ruler measurement, 
that is, the co-moving sound horizon scale at the end of drag epoch, $r_{\rm drag} = r_{\rm d}$. 
Assuming the flat $\Lambda$CDM cosmology, analyses of the CMB measurements from 
the Planck collaboration~\cite{Planck2018} and the WMAP team~\cite{wmap9} give 
$r_{\rm d} = 147.09 \pm 0.26$ Mpc and 
$r_{\rm d} = 152.99 \pm 0.97$ 
Mpc~\footnote{\url{https://lambda.gsfc.nasa.gov/product/map/dr5/params/%
lcdm_wmap9_spt_act.cfm}}, respectively. 
But there are also estimates of the sound horizon scale at low redshift combining data from 
large-scale structure: $r_{\rm d} = 150.0 \pm  4.7$ Mpc (CSB), 
$r_{\rm d} = 143.9 \pm 3.1$ Mpc (CSBH), where C-S-B-H indicate a combination of data 
from Cosmic Chronometers, SNe, BAO data, and local $H_0$ measurement (for details, 
see~\cite{rd_03}). 
An interesting information regarding the estimate of $r_{\rm d}$ using CMB data is that this 
derivation can be somehow biased by model hypotheses~\cite{Sutherland}. 
For this, the literature exhibits the efforts to obtain a model-independent estimate of 
$r_{\rm d}$~\cite{rd_03,Bernal}. 
An estimate of this type obtains $r_{\rm d} = 136.7 \pm 4.1$ Mpc~\cite{Bernal}, which is in 
tension of $\sim 2.5\,\sigma$ and $\sim 3.8\,\sigma$ with the Planck and WMAP values, 
respectively (for other analyses see, e.g.,~\cite{Sutherland,deCarvalho,theta_BAO_data}). Recently, final measurements  from the completed SDSS lineage of experiments in large-scale structure  provide $r_{\rm d} = 149.3 \pm 2.8$ Mpc \cite{SDSS_final}, 
in good agreement with Planck-CMB estimate.

The main aim of this work is to obtain constraints on some cosmological parameters of interest 
in current literature, namely $H_0$ and $r_{\rm d}$, independent of CMB and local 
distance ladder data, using sets of data obtained following  weakly model-dependent or model-independent 
approaches. 
Our analyses are done in $\Lambda$CDM/o$\Lambda$CDM models for comparison 
because the $H_0$ tension reported in the literature has been obtained within these models.
Such analyses are important, and of great interest,  providing an alternative 
way to investigate the current observational tension on these parameters and whose results 
can shed light on this problem. 
To achieve these objectives, in this work we use measurements of the transversal BAO scale 
($\theta_{\rm BAO}$), data obtained following an approach that weakly depends on the 
assumption of a cosmological model, as described in ref.~\cite{Sanchez11} 
(all these measurements were obtained following the same methodological approach, 
however, since the clustering analyses were performed with diverse cosmological tracers 
--blue galaxies, luminous red galaxies, quasars-- one should be careful with the systematics 
of each dataset~\cite[for some tests to deal with systematics in data analyses see, 
e.g.,][]{BMRT,deCarvalho20,
Marques20}). 
A distinctive feature of these data is that they were measured without assuming a geometry of the Universe; this is a crucial advantage as compared with data sets  obtained under the hypothesis of a flat geometry, an attribute that may bias cosmological parameter analyses.

For recent discussions on the cosmological constraints investigations under the perspective of BAO measurements by other groups see, e.g.,  \cite{baot1,baot2,baot3,baot4}.

In these combined analyses we also use the big bang nucleosynthesis (BBN) data, information 
from gravitationally lensed quasars with measured time delays (H0LiCOW data), and 
the cosmic chronometers (CC) data. 
We found that a robust analysis from these data sets is possible to get an accuracy up to 
$\sim$1.7\% on $H_0$, and this parameter lives in the range to be compatible with local 
measurements of $H_0$. 
To the authors' knowledge, this is the first $H_0$ measurement using BAO data information 
plus others data sets obtained in a weakly model-dependent way, able to generate high $H_0$ 
values, in order to be compatible with local and model-independent measures, within the 
$\Lambda$CDM framework. 

The paper is structured as follows. In the next section we present the data sets used in this work 
and the statistical methodology. In section \ref{Results} we discussed the main results of our 
analysis. In section \ref{Final_remarks} we outline our final considerations and perspectives.

\section{Methodology}
\label{Methodology}
We describe below the observational data sets and the statistical methods that we use to explore our parameter space. 
\\

\begin{figure*}
\begin{center}
\includegraphics[width=3.in]{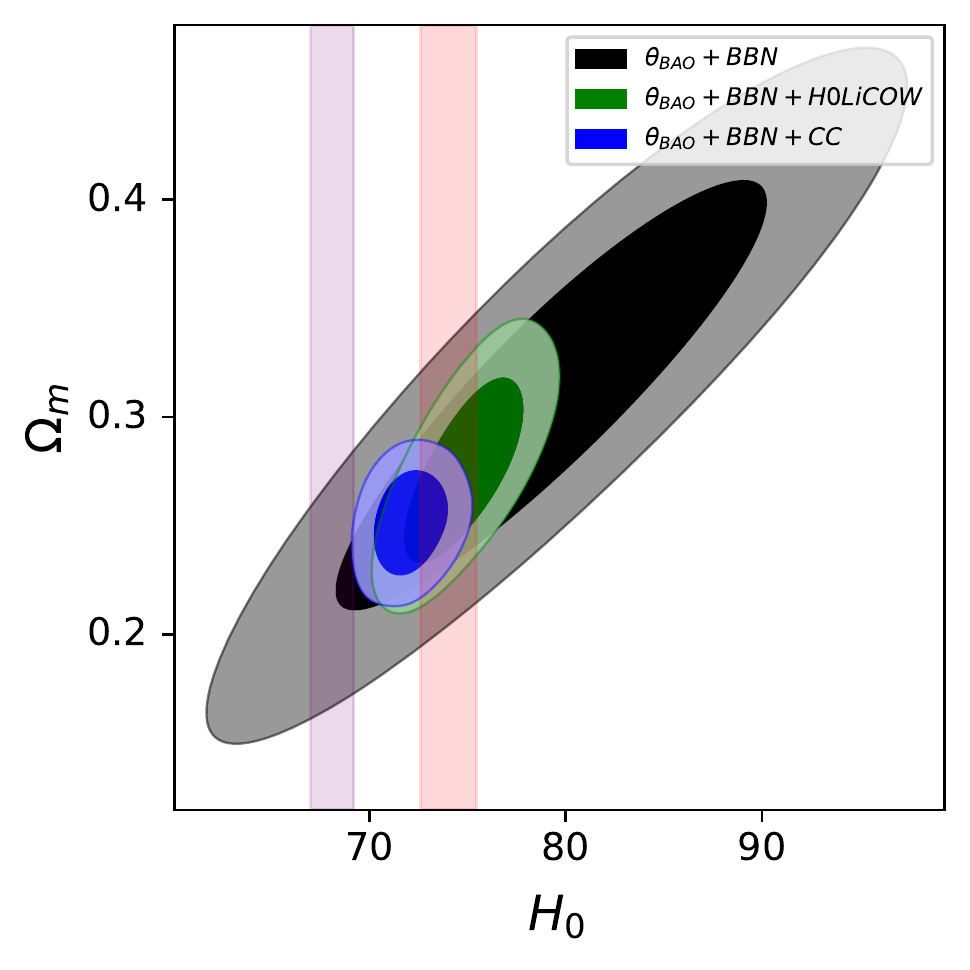} \,\,\,\,
\includegraphics[width=3.in]{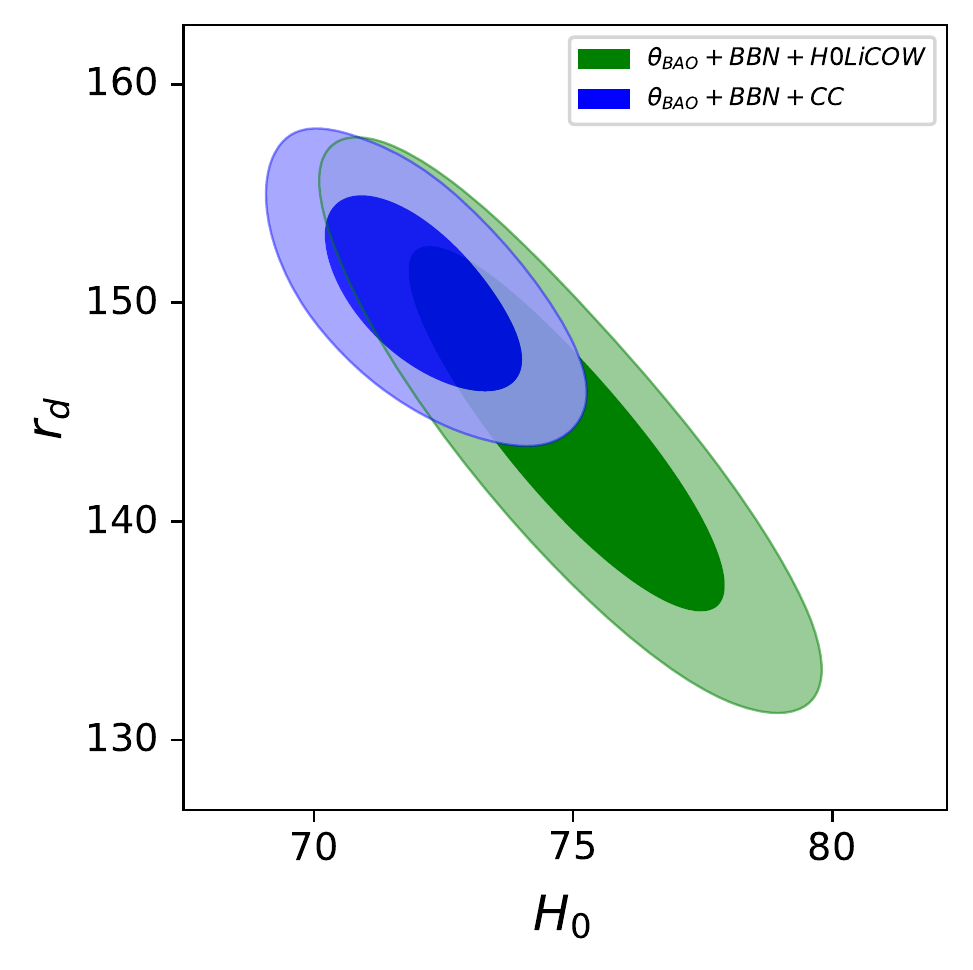}
\caption{Left panel: The 68\% CL and 95\% CL regions in the $H_0 - \Omega_m$ plane, inferred from $\theta_{BAO}$ + BBN analyses in combination with H0LiCOW and CC data. 
The vertical light-purple and light-red bands correspond to $H_0$ from BAO + BBN taken from  \cite{BAO_BBN_01} and the SHOES measurement \cite{R19}, respectively. 
Right panel: The 68\% CL and 95\% CL regions in the $H_0 - r_{\rm d}$ plane from 
$\theta_{BAO}$ + BBN + H0LiCOW and $\theta_{BAO}$ + BBN + CC analyses. 
The parameter $H_0$ is measured in units of km s${}^{-1}$ Mpc${}^{-1}$ and $r_{\rm d}$ 
in Mpc.}
\label{fig:1}
\end{center}
\end{figure*}

\textbf{Transversal BAO:} Let us adopt 15 BAO measurements, $\theta_{\mbox{\sc bao}}(z)$, 
obtained in a weakly model-dependent approach, compiled in table I in \cite{theta_BAO_data}. 
These measurements were obtained using public data releases (DR) of the Sloan Digital Sky 
Survey (SDSS), namely: DR7, DR10, DR11, DR12, DR12Q (quasars)~\cite{SDSS}. 
It is important to notice that due to the cosmological-independent methodology used to perform 
these transversal BAO measurements their errors are larger than the errors obtained using 
a fiducial cosmology approach. 
The reason for this fact is that, while in the former methodology the error is given by the 
measure of how large is the BAO bump, in the latter approach the model-dependent best-fit 
of the BAO signal quantifies a smaller error. 
Typically, in the former methodology the error can be of the order of $\sim 10\%$, but in 
some cases it can arrive to 18\%, and in the later approach it is of the order of few 
percent~\cite{Sanchez11}. 

Another important feature of these transversal BAO data is that the data points are not  correlated. 
In fact, the methodology adopted to perform these measurements excluded the  possibility for covariance between data points because the analyses of the 2-point angular correlation function were done with cosmic objects belonging to disjoint redshift bins 
(i.e., bins that are not overlapped and moreover separated by a minimum $\delta_z$, 
separation that excludes the possibility that errors in the redshifts could put one cosmic 
object in a contiguous bin).\\

\textbf{BBN:} The deuterium abundance and the radiative capture of protons on deuterium 
to produce $3He$ is one the most widely used primordial elements for constraining the 
baryon density. 
The empirical value for the reaction rate is computed in \cite{BBN}, constraining the baryon 
density to $100\,\Omega_b h^2 =  2.235 \pm 0.016$, where the dimensionless parameter 
$h \equiv H_0/$(100 km/s/Mpc) is the reduced Hubble constant. 
We adopt this value of $100\,\Omega_b h^2$ as a Gaussian prior likelihood in our analyses.
\\

\textbf{H0LiCOW:}
A powerful geometric method to measure $H_0$ is offered by the gravitational lensing. The time delay between multiple images, produced by a massive object (lens) and the gravitational potential between a light-emitting source and an observer, can be measured by looking for f\/lux variations that correspond to the same source event. This time delay depends on the mass distribution along the line of sight and in the lensing object, and it represents a complementary and independent approach with respect to the CMB and the distance ladder. Due to their variability and brightness, lensed quasars have been widely used to determine $H_0$ (see, e.g., \cite{Sereno14,Kumar15,Bonvin17} and references therein).
The time delay is highly sensitive to $H_0$, but with a weak dependence on other cosmological parameters. In the present work, we use the six systems of strongly lensed quasars reported by the H0LiCOW Collaboration \cite{H0LiCOW}, its time delay distances, to constraint  directly all free parameters in our baseline.
\\

\textbf{CC:} The late expansion history of the Universe can be studied in a model-independent fashion by measuring the age difference of cosmic chronometers (CC), such as old and passively evolving galaxies that act as standard clocks \cite{Jimenez02, CC}. 
In our analysis we consider the measurements of CC as presented in \cite{CC}. 
\\

We ran \texttt{CLASS}+\texttt{MontePython} code \cite{class,MP1,MP2} using Metropolis-Hastings mode to derive constraints on cosmological parameters from the BAO+BBN, BAO+BBN+H0LiCOW and BAO+BBN+CC data combination. Our baseline parameters are  $100 \omega_b \in [0.8 , 2.4]$, $\omega_{cdm} \in [0.01 , 0.99]$, $H_0 \in [10, 100]$, and $\Omega_k \in [-1, 1]$. The parameters $\omega_b \equiv \Omega_b h^2$ and $\omega_{cdm} \equiv \Omega_{cdm} h^2$ are the baryon and the cold dark matter energy densities, respectively. The parameter $H_0$ is the Hubble constant and $\Omega_k$ the spatial curvature. 

In a first round of analysis we consider that the  background expansion framework is fix assuming a flat-$\Lambda$CDM scenario. Next, we also analyze the case $\Lambda$CDM + $\Omega_k$. All of our runs reached a Gelman-Rubin convergence criterion of $R-1 < 10^{-3}$. In what follows, we discuss the main results of our analyses.
\\

\section{Results}
\label{Results}

The left panel of figure~\ref{fig:1} shows the parametric space in the plane $H_0-\Omega_m$ 
from $\theta_{BAO}$+BBN, $\theta_{BAO}$+BBN+H0LiCOW and $\theta_{BAO}$+BBN+CC 
data combination. We find $H_0  = 74.88_{-2.1}^{+1.9}$ km s${}^{-1}$ Mpc${}^{-1}$ and 
$H_0 = 72.06_{-1.3}^{+1.2}$ km s${}^{-1}$ Mpc${}^{-1}$ at 68\% confidence level (CL) from 
$\theta_{BAO}$+BBN+H0LiCOW and $\theta_{BAO}$+BBN+CC, respectively. 
The total matter density (baryon + dark matter density) is fit to be $\Omega_{m} = 0.2763_{-0.028}^{+0.027}$ and $\Omega_{m} =0.2515_{-0.016}^{+0.016}$ at 68\% CL from $\theta_{BAO}$+BBN+H0LiCOW and $\theta_{BAO}$+BBN+CC, respectively. Since the measurements of $\theta_{BAO}$ have error bars a bit larger than other BAO data compilations, one can notice that the $H_0$ parameter becomes more degenerate from $\theta_{BAO}$ + BBN constraints when compared to other BAO + BBN analyses performed in the literature \cite{BAO_BBN_01,BAO_BBN_02}. Interesting to note that the $H_0-\Omega_m$ plane, from $\theta_{BAO}$ data, also tends to be positively correlated, but generating high $H_0$ values. We add H0LiCOW lenses and CC data to better bounds the parameter space. In the Appendix we show the consistency between these data sets. In Figure~\ref{fig:1}, the horizontal light purple and light red bands correspond 
to $H_0$ values from the BAO + BBN analysis ~\cite{BAO_BBN_01} and the SHOES measurement ~\cite{R19}, respectively. We note that $H_0$ is at $\sim$2$\sigma$ and $\sim$2.5$\sigma$ tension from $\theta_{BAO}$+BBN+H0LiCOW and $\theta_{BAO}$+BBN+CC, respectively, when compared to the measurements performed in \cite{BAO_BBN_01}. In contrast, our $H_0$ estimates are in agreement with SHOES~\cite{R19}. 

Therefore, combining $\theta_{BAO}$ with other data obtained in a model-independent way, 
and without using CMB and supernovae data, we see their concordance with local 
measurements of $H_0$. 
A direct interpretation of why the $\Lambda$CDM scenario is generating high $H_0$ values, is 
because our global fit predicts less dark matter today --in  contrast, more dark energy-- 
via the relation $\Omega_{m} + \Omega_{DE} = 1$, where $\Omega_{m} = \Omega_{b} + 
\Omega_{DM}$. Notice that $\Omega_{b}$ here is determined from BBN information. 
So, the change on $\Omega_{m}$ estimate is due to dark matter density only, once the 
radiation (photons + neutrinos) contribution is negligible at low-$z$. Because our joint analysis  predicts more 
dark energy at late times, the Universe expands faster, generating a larger $H(z)$ evolution 
and high $H_0$ values. In \cite{theta_BAO_data}, was analyzed CMB 
+ $\theta_{BAO}$, where we report $H_0 = 69.23 \pm 0.50$ km s${}^{-1}$ Mpc${}^{-1}$, where 
we can see a displacement of $\sim + 2$ km s${}^{-1}$ Mpc${}^{-1}$, in relation to the Planck + 
BAO analysis made by the Planck Collaboration ~\cite{Planck2018}. 
Again, it is clear that $\theta_{BAO}$ tends to generate higher $H_0$ values in comparison with 
other BAO compilation in literature. 
The $H_0$ value from CMB data is inferred analyzing the first acoustic peak position, which 
depends on the angular scale $\theta_{*} = d_s^{*}/D_A^{*}$, where $d_s^{*}$ is the sound 
horizon at decoupling (the distance a sound wave traveled from the big bang to the epoch of 
the CMB-baryons decoupling) and $D_A^{*}$ is the angular diameter distance at decoupling, 
which in turn depends on the expansion history, $H(z)$, after decoupling, controlled also by 
the ratio $\Omega_{DM}/\Omega_{DE}$ and $H_0$ mainly. Our joint fit is generating a larger 
$H(z)$ and, at the same time, changing the slope of the Sachs-Wolfe plateau, that is, the 
late-time integrated Sachs-Wolfe effect (ISW). Thus, our joint fit (CMB + $\theta_{BAO}$) is 
changing primarily the $D_A^{*}$ history, increasing the angular diameter distance to the 
last scattering surface, thus generating high estimates on the $H_0$ parameter.

\begin{figure}
\begin{center}
\includegraphics[width=3.in]{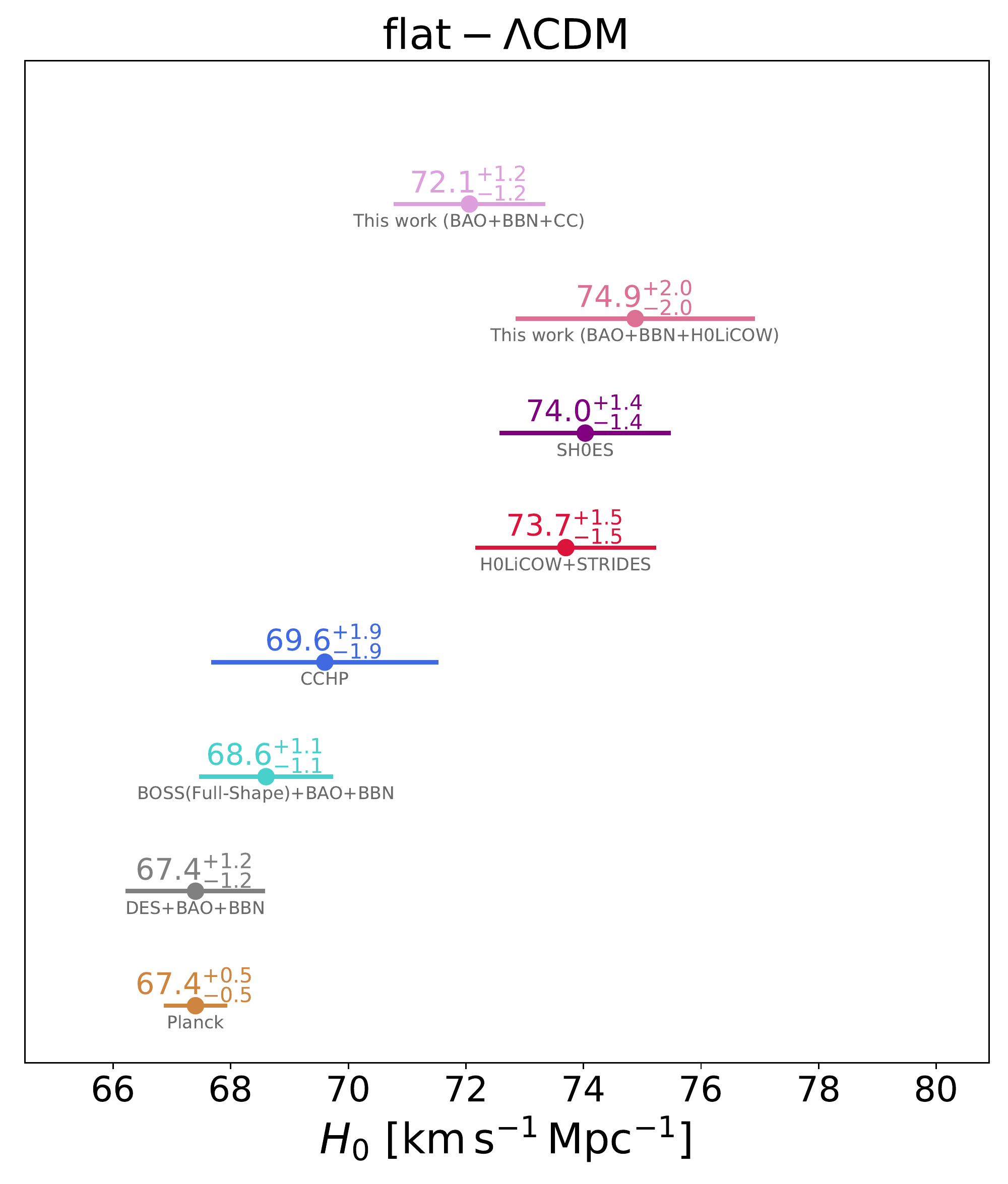}
\caption{Compilation of $H_0$ measurements taken from recent literature, namely, 
from Planck collaboration (Planck) \cite{Planck2018}, Dark Energy Survey Year 1 Results (DES+BAO+BBN) \cite{DES}, the final data release of the BOSS data (BOSS Full-Shape+BAO+BBN) \cite{BOSS_H0}, The Carnegie-Chicago Hubble Program (CCHP) \cite{Freedman}, H0LiCOW collaboration (H0LiCOW+STRIDES) \cite{H0LiCOW}, SH0ES \cite{R19}, in direct comparison 
with the $H_0$ constraints obtained in this work from $\theta_{BAO}$+BBN+H0LiCOW and 
$\theta_{BAO}$+BBN+CC analyses within the flat-$\Lambda$CDM scenario.}
\label{fig:H0_tension}
\end{center}
\end{figure}

Figure \ref{fig:H0_tension} shows a compilation of $H_0$ measurements taken from the recent literature for direct comparison with our results. We can notice that $H_0$ obtained in this work is in agreement with SH0ES, H0LiCOW+STRIDES and CCHP. Our estimates start to have a significant tension when compared to measures involving other BAO data compilation and Planck data only.

The right panel of Figure~\ref{fig:1} 
shows the parametric space in the $H_0-r_{\rm d}$ plane. 
We find $r_{\rm d}=144.1_{-5.5}^{+5.3}$ Mpc (from $\theta_{BAO}$+BBN+H0LiCOW) and $r_{\rm d} =150.4_{-3.3}^{+2.7}$ Mpc (from $\theta_{BAO}$+BBN+CC) at 68\% CL. Both measures are compatible with each other. This fit represents an $r_{\rm d}$ constraint  obtained independently of CMB data. For a qualitative comparison, the Planck team reported the value $r_{\rm d}=147.21 \pm 0.23$ Mpc from CMB + BAO joint analysis. We see that this estimate is in concordance with ours. 
Regarding analyses independent of the CMB data, we can mention, for instance, a model-independent reconstruction of $H(z)$ done in ref.~\cite{rd_01},  where it is reported 
$r_{\rm d}=148.48_{-3.74}^{+3.73} \pm 0.23$ Mpc. 
In ref.~\cite{rd_02}, the sound horizon at radiation drag is considered as a standard ruler, and it is  found $r_{\rm d}= 142.8 \pm 3.7$ Mpc. Also, in 
ref.~\cite{rd_03} the authors found $r_{\rm d}= 143.9 \pm 3.1$ Mpc using CC, SNe~Ia, BAO, and a local measurement of $H_0$. Using the inverse distance ladder method, the DES collaboration found $r_{\rm d}= 145.2 \pm 18.5$ Mpc from SNe Ia and BAO measurements \cite{rd_04}. Our estimates are consistent with these measurements too. We note that only the $r_{\rm d}$ from $\theta_{BAO}$+BBN+CC joint analysis is in $\sim$1$\sigma$ tension with 
ref.~\cite{rd_03}. 
Other results independent of CMB data were obtained in \cite{rd_05,rd_06,rd_07}. 

\begin{figure*}
\begin{center}
\includegraphics[width=3.in]{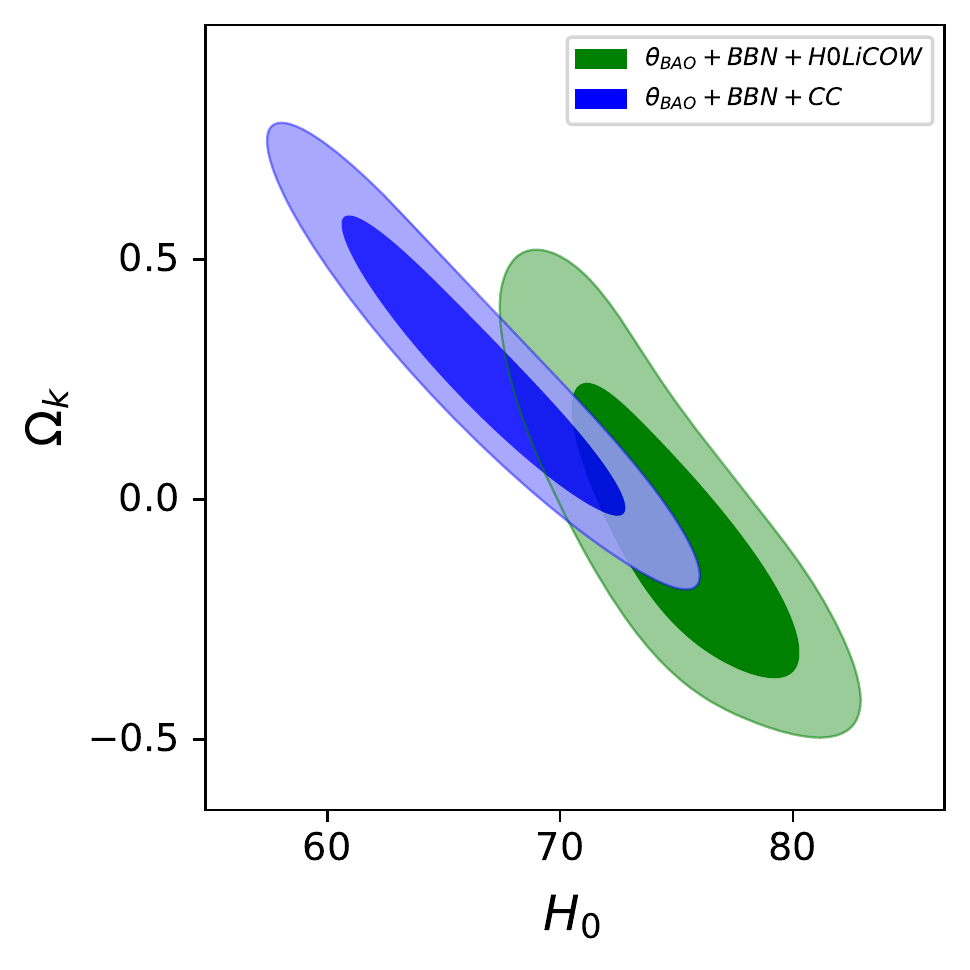} \,\,\,\,
\includegraphics[width=3.in]{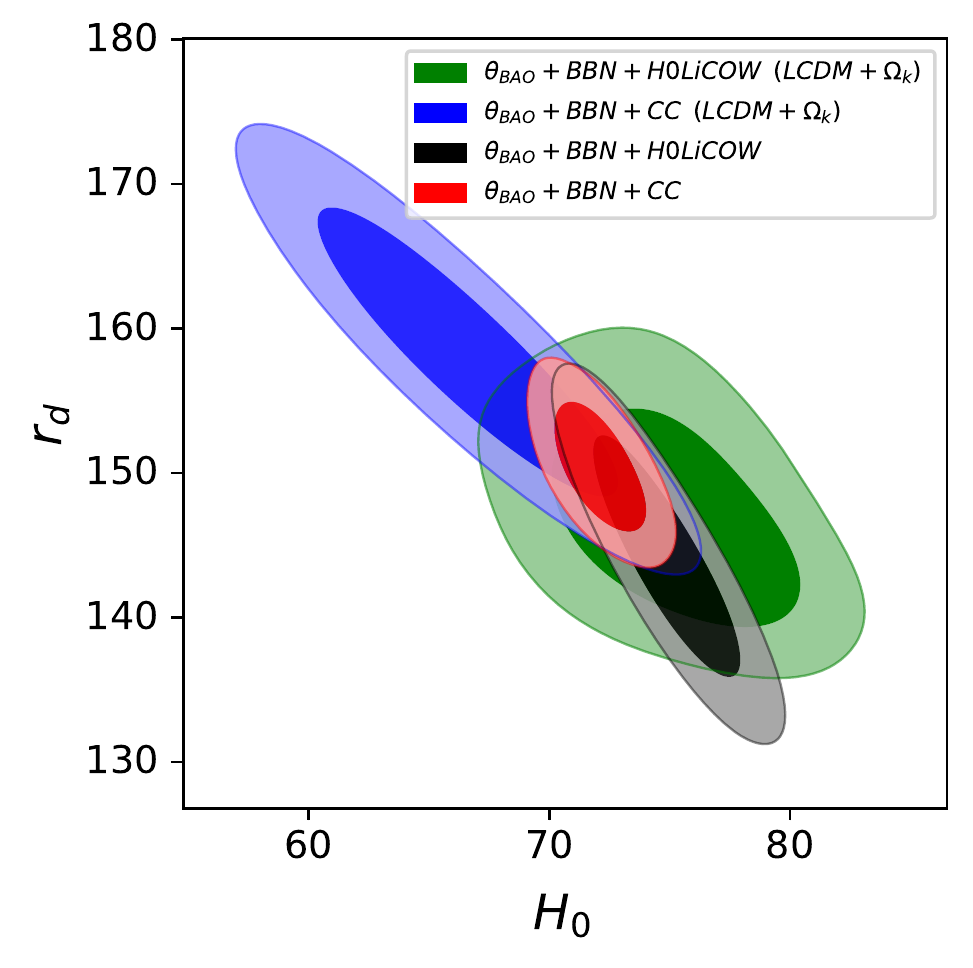}
\caption{Left Panel: The 68\% CL and 95\% CL regions in the $H_0 - \Omega_k$ plane inferred 
from $\theta_{BAO}$ + BBN analyses in combination with H0LiCOW and CC data. 
Right Panel: Parametric space in the $H_0 - r_{\rm d}$  plane from the analyses with and without 
the $\Omega_k$ parameter. 
The $H_0$ parameter is measured in units of km s${}^{-1}$ Mpc${}^{-1}$ and $r_{\rm d}$ in Mpc.}
\label{fig:H0_Omegak}
\end{center}
\end{figure*}

\subsection{Adding spatial curvature}

Until now we have performed statistical analyses considering the flat $\Lambda$CDM model. 
Here we extend the parameter space to analyse these important quantities, $H_0$ and 
$r_{\rm d}$, within a model beyond the flat $\Lambda$CDM. 
For this, we now consider the spatial curvature as a free parameter, i.e., $\Omega_{k} \ne 0$. 
As we shall see below, our analyses show compatibility with $\Omega_k = 0$, within 
$1\,\sigma$ error, although we observe an enlargement of the error bars (as expected 
because there is one more parameter in the analysis).

Analyzing $\Lambda$CDM + $\Omega_{k}$ from BAO+BBN+H0LiCOW we find: 
$H_0 = 75.08_{-3.0}^{+3.5}$ km s${}^{-1}$ Mpc${}^{-1}$ and 
$\Omega_{k}= -0.0697_{-0.26}^{+0.14}$. 
As argued in \cite{H0LiCOW}, the time delay is highly sensitive to $H_0$, but with a weak 
dependence on other parameters. 
Thus, we can note that when assuming $\Omega_{k}$ as a free parameter, and considering the 
H0LiCOW sample, no significant changes are observed in the baseline of parameters. 
Only the effect of slightly increasing the error bars due to the presence of an extra parameter, 
$\Omega_{k}$. 
This scenario can change the perspectives when considering BAO+BBN+CC; in fact, in this case we find $H_0 = 66.54 \pm 3.76$ km s${}^{-1}$ Mpc${}^{-1}$ and $\Omega_{k} = 0.2764_{-0.28}^{+0.17}$. In this case, considering $\Omega_{k}$ as a free parameter, this can significantly changes the evolution of the $H(z)$ function, which depends directly on all physical species and geometrical effects. 
We note this effect by observing an enlargement and shift in the estimate and error bar of 
$H_0$ to accommodate $\Omega_{k}$ effects into the $H(z)$ function. 
In this particular case we find $\Omega_{m} = 0.2378_{-0.019}^{+0.02}$. 
Thus, to accommodate $\Omega_{k}$ effects, looking through the relationship $\Omega_{k}+\Omega_{m }+\Omega_{\Lambda} = 1$, and using for comparison the $\Omega_{m}$ best fit derived in the previous 
section without $\Omega_{k}$, we note that the  presence of $\Omega_{k}$ decreases mainly the value of $\Omega_{\Lambda}$. 
The left panel in Figure \ref{fig:H0_Omegak} shows the constraints in 
the plane $H_0 - \Omega_k$. 
We did not find significant deviations from the $\Omega_k = 0$ case.  The curvature parameter $\Omega_k$ have been discussed through other observations recently in \cite{Omega_K_01,Omega_K_02,Omega_K_03,Omega_K_04,Omega_K_05}.

The right panel of figure \ref{fig:H0_Omegak} shows the 68\% CL and 95\% CL regions in the 
$H_0 - r_{\rm d}$ plane, analyses done with and  without the  parameter $\Omega_k$ for comparison. 
Assuming $\Lambda$CDM + $\Omega_{k}$, we find  $r_{\rm d}  =146.9_{-5.7}^{+4.2}$ Mpc 
(from BAO+BBN+H0LiCOW) and  $r_{\rm d}=158.3_{-7.1}^{+5.8}$ Mpc (from BAO+BBN+CC). 
As previously commented, in the BAO+BBN+H0LiCOW joint analyses no significant deviations were observed, as compared to the flat case. 
In the BAO+BBN+CC analyses, we can clearly notice an enlargement for higher 
values in $r_{\rm d}$, possibly due to the change in $H_0$ and the strong correlation of 
$r_{\rm d}$ with $H_0$. 
It is important to emphasize that all analyses  investigated here agree with each other. 

\subsection{SDSS final release}

\begin{figure}
\begin{center}
\includegraphics[width=3.in]{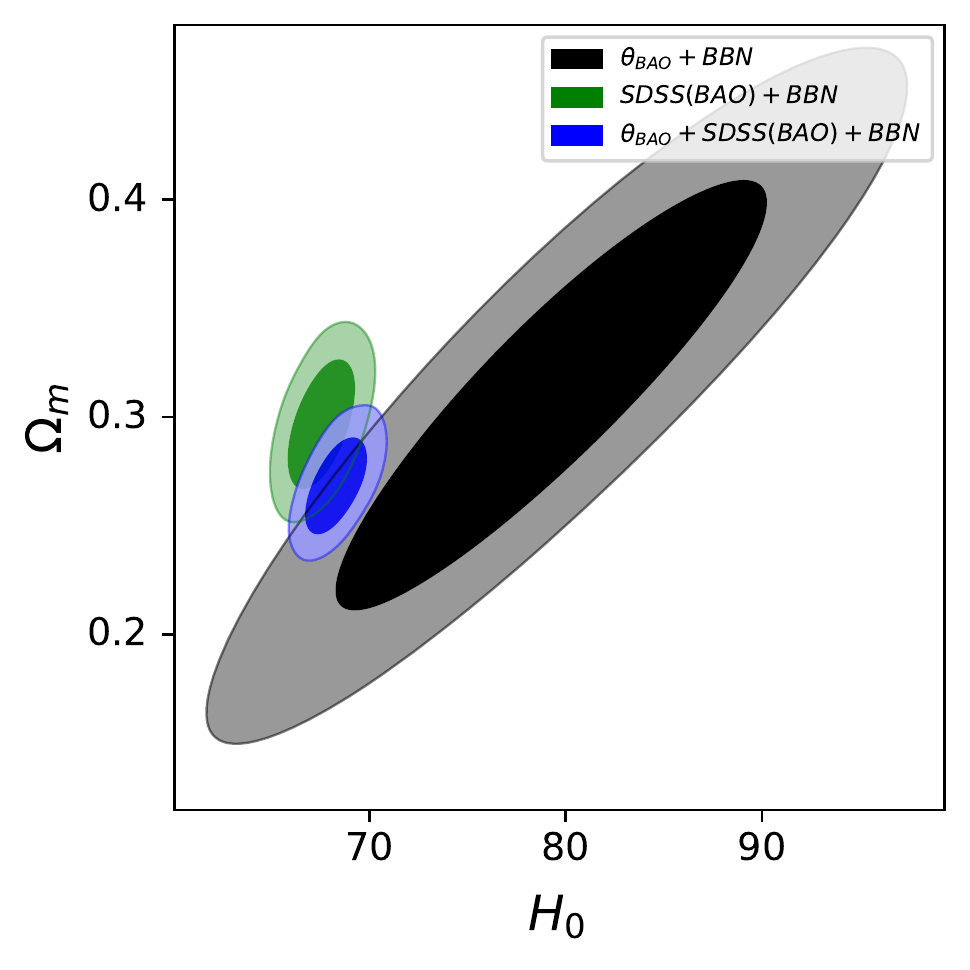} 
\caption{Parametric space in the $H_0 - \Omega_m$ plane inferred from $\theta_{BAO}$ + BBN, SDSS (BAO) final release + BBN, and $\theta_{BAO}$ + SDSS (BAO) final release + BBN joint analyses. The $H_0$ parameter is measured in units of km s${}^{-1}$ Mpc${}^{-1}$.}
\label{fig:BAO_SDSS}
\end{center}
\end{figure}

During the final stage of preparation of this work, the SDSS collaboration released their BAO final measurements covering eight distinct redshift intervals, obtained and improved over the past 20 years \cite{SDSS_final}. Given the importance of these data for cosmology in recent years, here we perform a brief analysis for comparison with our measurements, using the $D_V(z)/r_d$, $D_M(z)/r_d$, and $D_H(z)/r_d$ measurements compiled in Table 3 in \cite{SDSS_final}, regarding BAO-only data. In what follows, we call this data compilation by SDSS (BAO). We assume that the uncertainties are Gaussian approximations to the likelihoods for each tracer ignoring the correlations between measurements (as suggested in the SDSS collaboration paper).

Figure \ref{fig:BAO_SDSS} shows the 68\% CL and 95\% CL regions in the 
$H_0 - \Omega_m$ plane from $\theta_{BAO}$ + BBN, SDSS (BAO) + BBN, and  $\theta_{BAO}$ + SDSS (BAO) + BBN joint analysis. 
Evidently, the accumulation of accuracy and  improvement in the measurements over the years make  the analysis of SDSS (BAO) + BBN very robust in the  errors determination, in a direct comparison with $\theta_{BAO}$ + BBN only (see the figure~\ref{fig:BAO_SDSS}). 
We find $H_0 = 68.32_{-1.1}^{+0.98}$ km s${}^{-1}$ Mpc${}^{-1}$, $r_{d} = 151.9_{-2.8}^{+3}$ Mpc, $\Omega_{m}=0.27_{-0.016}^{+0.015}$ at 68\% CL from $\theta_{BAO}$ + SDSS (BAO) + BBN joint analyses. This estimate of $H_0$, influenced by SDSS (BAO) data, is in agreement with the Planck-CMB data, and in  $\sim$4$\sigma$ tension with the SHOES~\cite{R19} value. 
There is no tension on the $r_{d}$ parameter when  compared to Planck-CMB data.

We check the individual (in)consistency between BBN+$\theta_{BAO}$ and BBN + SDSS (BAO) constraints. The $H_0$ values obtained separately with these data sets differ in $\sim$2$\sigma$, and the $\Omega_m$ parameter is in full statistical compatibility one to each other.

\section{Final remarks}
\label{Final_remarks}

We obtained accurate constraints on $H_0$ and $r_{\rm d}$ parameters, independently of 
CMB data and local distance ladder data. 
In this work we are motivated to look how recent  transversal BAO measurements (that is, from  $\theta_{\rm BAO}$ estimates~\cite{theta_BAO_data}), in combination with other  model-independent data sets, can bound these  parameters and what direction do they take in light of recent observational tensions, especially in the context of the $H_0$ tension. We find an accuracy of $\sim$2.6\% and $\sim$1.7\% on $H_0$ from $\theta_{BAO}$+BBN+H0LiCOW and $\theta_{BAO}$+BBN+CC, respectively. We observe that both values are  compatible with local estimates of $H_0$, and in tension with Planck data only and some joint analyses in  combination with other BAO compilations of the literature. Our results show that it is possible to use a robust compilation of BAO data, i.e., the  $\theta_{BAO}$ compilation, in such a way that the tension on the $H_0$ parameter is minimized or even not exist, when compared to local and 
model-independent measurements.

An interesting perspective regards the measurements of transversal BAO data. 
With arriving new data from ongoing astronomical surveys we expect new transversal BAO 
measurements, with both features: more precise estimates and performed at diverse redshifts. 
In fact, these data has shown potential to constrain better $r_{\rm d}$ and $H_0$, important 
quantities in modern cosmology because they provide absolute scales to measure the Universe 
evolution at opposite sides. 

\begin{acknowledgments}
\noindent 
RCN would like to thank the agency FAPESP for financial support under the project 
No. 2018/18036-5. AB acknowledges a CNPq fellowship.
\end{acknowledgments}

\section*{Consistency between $\theta_{BAO}$, CC, and H0LiCOW samples}
\label{consistency_test}

\begin{figure}
\begin{center}
\includegraphics[width=3.in]{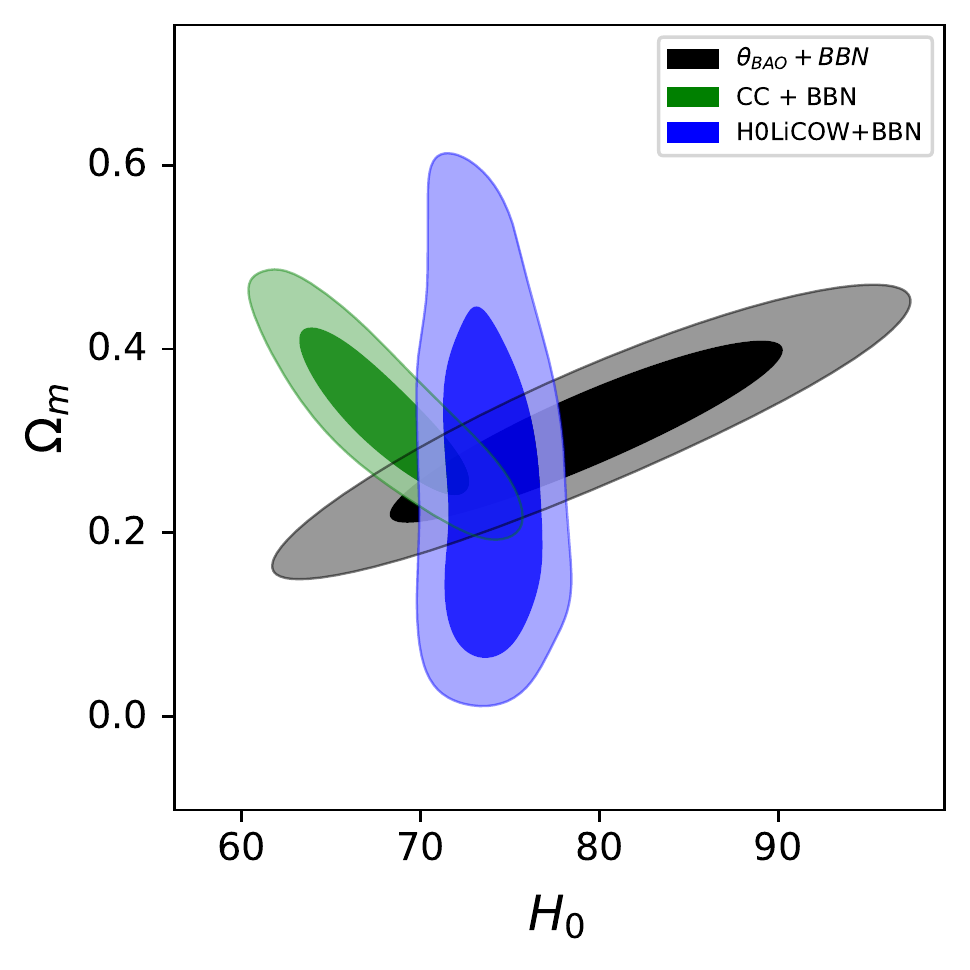} 
\caption{Parametric space in the $H_0 - \Omega_m$ plane inferred from $\theta_{BAO}$ + BBN, CC + BBN and H0LiCOW+BBN. The $H_0$ parameter is measured in units of km s${}^{-1}$ Mpc${}^{-1}$.}
\label{fig:bao_consistency}
\end{center}
\end{figure}

The aim of this section is to show that the $\theta_{BAO}$, CC, and H0LiCOW data sets are consistent with each other. These analyses are  important because make clear for the readers that these data sets can be combined properly without worrying about any possible inconsistency and/or tensions between them. Thus, these data can be used in joint analyses to test phenomenological  models and hypotheses in cosmology.

Figure \ref{fig:bao_consistency} shows the  parametric space in the $H_0 - \Omega_m$ plane inferred from $\theta_{BAO}$ + BBN, CC + BBN, and BBN+H0LiCOW. The BBN information is taken for improve any possible degeneracy in the total matter density. We can conclude that these three data sets, i.e., $\theta_{BAO}$, CC, and H0LiCOW samples, are fully consistent with each other at $\lesssim$ 1$\sigma$.

\end{document}